\documentstyle[12pt]{JHEP3}
\newcommand{\be}[1]{ \begin{equation}\label{#1} }
\newcommand{\ee}{\end{equation}}
\newcommand{\bea}[1]{\begin{eqnarray}\label{#1} }
\newcommand{\eea}{\end{eqnarray}}
\newcommand{\eq}[1]{(\ref{#1})}
\def\ZZZ{{\hskip-3pt\hbox{ Z\kern-1.6mm Z}}}
\def\zzz{{\hskip-3pt\hbox{ z\kern-1mm z}}}

\newcommand{\vp}{\varphi}

\newcommand{\vptl}{\tilde\varphi}

\newcommand{\half}{{1\over 2}}

\def\one{{\hbox{ 1\kern-.8mm l}}}
\def\zero{{\hbox{ 0\kern-1.5mm 0}}}

\title{Quantum ${\cal W}$-symmetry in AdS$_3$}

\author{
Matthias R.\ Gaberdiel$^a$, Rajesh Gopakumar$^b$ and Arunabha Saha$^b$ \\ 
$^a$Institut f\"ur Theoretische Physik, ETH Zurich, \\
$\;$CH-8093 Z\"urich, Switzerland \\
$\;$\email{gaberdiel@itp.phys.ethz.ch}\\ \\ 
$^b$Harish-Chandra Research Institute, \\
$\;$Chhatnag Road, Jhusi,\\
$\;$Allahabad, India 211019\\
$\;$\email{gopakumr@hri.res.in, arunabha@hri.res.in}}

\abstract{It has recently been argued that, classically, massless higher spin theories in AdS$_3$ have an 
enlarged  ${\cal W}_N$-symmetry as the algebra of asymptotic isometries. In this note we provide 
evidence that this symmetry is realised (perturbatively) in the quantum theory. We perform a 
one loop computation of the fluctuations for a massless spin $s$ field around a thermal AdS$_3$ background. 
The resulting determinants are evaluated using the heat kernel techniques of arXiv:0911.5085.
The answer factorises holomorphically, and the contributions from the various spin $s$ fields organise
themselves into vacuum characters of the ${\cal W}_N$ symmetry. 
For the case of the $hs(1,1)$ theory consisting of an infinite tower of massless higher spin particles,
the resulting answer can be simply expressed in terms of (two copies of) the MacMahon function.}

\preprint{HRI/ST/1008}

\begin{document}

\section{Introduction}

One of the remarkable features of AdS spacetimes is the existence of interacting theories 
of massless particles with spin greater than  two \cite{Vasiliev:2003ev}. As is well known, 
it is impossible to have such theories in flat spacetime. Typically, the consistency of such theories in 
AdS spacetimes requires the introduction of an infinite tower of particles with arbitrarily 
high spin.
These theories, coupled to gravity, are thus in some sense intermediate between 
conventional field theories of gravity involving a finite number of fields on the one hand, and 
string theories on the other (see \cite{Bekaert:2005vh} for an introduction to these matters). 

This observation takes on added significance in the context of the 
AdS/CFT correspondence. Higher spin theories provide an opportunity to understand the 
correspondence beyond the (super)gravity limit without necessarily having the full string theory 
under control. In fact, there is the tantalising 
possibility of a consistent truncation to this subsector within a string theory which might, 
in itself, be dual to a field theory on the boundary of the AdS spacetime. A conjecture of precisely 
this nature is that of Klebanov and Polyakov for the dual description of the $O(N)$ vector model 
in $2+1$ dimensions in terms of a higher spin theory on AdS$_4$ \cite{Klebanov:2002ja, Sezgin:2002rt}. 
Recent calculations have  provided non-trivial, interesting evidence for this conjecture,
see for example \cite{Giombi:2009wh,Giombi:2010vg,Koch:2010cy}.

One of the main hurdles in the exploration of this topic has been that, even at the classical level, 
it is very difficult to perform calculations. In fact, the formulation of higher spin theories is 
typically quite complicated. Therefore the study of these theories, especially at the quantum level, 
is in its infancy. 

Recently, attention has been drawn \cite{ Henneaux:2010xg, Campoleoni:2010zq} to the case of 
three dimensions where the classical theories are relatively more tractable since it is consistent
to consider theories involving a finite number of higher spin fields \cite{Aragone:1983sz}. Here the massless 
higher spin fields do not have any propagating degrees of freedom (as we will review in Sec. 1.1),
just as in the case of pure gravity. Nevertheless, as we have learnt in the case of theories of gravity 
on AdS$_3$, there can be interesting quantum dynamics captured by a two dimensional 
conformal field theory on the boundary. Thus these theories can be a useful stepping stone 
towards studying 
such higher spin theories in  dimension greater than three. 

An important first step was taken in \cite{ Henneaux:2010xg, Campoleoni:2010zq} by
studying the asymptotic symmetries of these higher spin theories at the classical level. 
The analysis is the analogue of the Brown-Henneaux study in pure gravity on AdS$_3$ 
\cite{Brown:1986nw}. In the case of theories with massless higher spin particles (of spin 
$s=3,\ldots,N$) 
one finds that the algebra of asymptotic isometries is enlarged from that of two copies 
(left- and right-moving) of the Virasoro algebra, to two copies of the ${\cal W}_N$ algebra. Like in the 
Brown-Henneaux analysis  one finds a central extension of the algebra already at the 
classical level.
The central charge is the same  \cite{ Henneaux:2010xg, Campoleoni:2010zq} as that of 
Brown-Henneaux \cite{Brown:1986nw}, namely
\be{bhcent}
c={3\ell\over 2G_N}\ .
\ee

In this paper, we study these massless higher spin theories on AdS$_3$ at the quantum level. 
More specifically, we perform a one loop calculation of the quadratic fluctuations of the fields about a 
thermal AdS$_3$ background. This requires a careful accounting of the gauge degrees of freedom 
of these fields. In particular, we show that the partition function reduces to the ratio of two determinants. 
For a spin $s$ field these involve Laplacians for transverse traceless modes of helicity $\pm s$ as 
well as $\pm (s-1)$
\be{dets}
Z^{(s)}=\Biggl[ \det\left(-\Delta  + \frac{s(s-3)}{\ell^2}  \right)^{\rm TT}_{(s)}\Biggr]^{-\frac{1}{2}}  \, 
\Biggl[ \det\left(-\Delta + \frac{s(s-1)}{\ell^2} \right)^{\rm TT}_{(s-1)}\Biggr]^{\frac{1}{2}}  \ .
\ee
In \cite{David:2009xg} such determinants were explicitly evaluated, in a thermal background, 
for arbitrary spin $s$, using the group theoretic techniques of 
\cite{Higuchi:1986wu,Camporesi:1990wm}. By applying the results of \cite{David:2009xg} 
we find that the one loop answer  factorises neatly into left and right moving pieces 
\be{detevl}
Z^{(s)}= \prod_{n=s}^\infty {1\over |1-q^n|^2}\ ,
\ee
where $q=e^{i \tau}$ is the modular parameter of the boundary $T^2$ of the thermal 
background. This generalises the expression for the case of pure gravity $(s=2)$ 
\cite{Maloney:2007ud}, as explicitly 
checked in \cite{Giombi:2008vd}. The expression \eq{detevl}  is seen to be the contribution to the character 
for a generator of conformal dimension $s$. Combining the different fields of spin $s=3,\ldots,N$, together
with the corresponding expression for the spin two case, one obtains indeed the vacuum character of 
the ${\cal W}_N$ algebra.

A straightforward generalisation of an argument of Maloney and Witten \cite{Maloney:2007ud} 
can now be made to show that this expression (together with the classical contribution 
$(q\bar{q})^{-{c\over 24}}$) is one loop exact in perturbation theory. This is to be understood in a 
particular scheme where the Newton constant is suitably renormalised while keeping $c$ fixed. 

It is interesting that if we consider the Vasiliev higher spin theory with left and right copies of the 
$hs(1,1)$ higher spin algebras, then we find a vacuum character of the ${\cal W}_{\infty}$ algebra,
which can be written as 
\be{hs}
Z_{hs(1.1)}=\prod_{s=2}^{\infty}\prod_{n=s}^{\infty}{1\over |1-q^n|^2} 
=\prod_{n=1}^{\infty}|1-q^n|^2\times \prod_{n=1}^{\infty}{1\over |(1-q^n)^n|^2}\ .
\ee
It is interesting that the answer can be naturally expressed in terms of the so-called MacMahon function
\be{macm}
M(q)=\prod_{n=1}^{\infty} {1\over (1-q^n)^n} \ .
\ee
\medskip

The organisation of this paper is as follows. In the next subsection we review some 
basic features of the massless higher spin fields. We will find it useful to decompose the 
fields in terms of transverse traceless modes of various helicities which will enable us to 
count the physical and gauge degrees of freedom. 
In Sec.~2, we lay out the basics for the calculation of the quadratic fluctuations, correctly taking 
into account the redundancy from the gauge modes. In Sec.~3 we obtain the one loop answer for 
the spin three case in a brute force manner. Sec.~4 shows how the answer for a general spin can 
be carried out without having to do too much work. Sec.~5 uses the results of \cite{David:2009xg} 
to  evaluate the determinants \eq{dets} explicitly in a thermal AdS$_3$ background to obtain \eq{detevl}. 
Sec.~6 relates these expressions to the vacuum characters of ${\cal W}_N$. We also comment on the case of 
${\cal W}_{\infty}$ and the relation to the MacMahon function. Sec.~7 contains closing remarks while  
the appendices describe our conventions and spell out some of the details of the spin three 
calculation.

\subsection{Counting Degrees of Freedom} 

Let us first recall some basic features of massless higher spin theories at the non-interacting level
\cite{Fronsdal:1978rb,Vasiliev:1980as} (see for example 
\cite{Campoleoni:2009je}  for a review and more references). 
The massless spin $s$ fields in three dimensions are completely symmetric tensors 
$\varphi_{\mu_1\mu_2\ldots \mu_s}$ subject to a double trace constraint
\be{dbtr}
\varphi_{\mu_5\ldots \mu_s \alpha\lambda}{}^{\alpha \lambda} \ =0.
\ee
This constraint only makes sense if  $s\geq 4$.  In, addition we have a gauge invariance leading to the 
identification of field configurations
\be{ginv}
\varphi_{\mu_1\mu_2\ldots \mu_s} \sim \varphi_{\mu_1\mu_2\ldots \mu_s}
+\nabla_{(\mu_1}\xi_{\mu_2 \ldots \mu_{s})}\ .
\ee
The gauge parameter $\xi_{\mu_2 \ldots \mu_{s}}$ is a symmetric tensor of rank $(s-1)$ which is, 
in addition, traceless, {\it i.e.}\  $\xi_{\mu_3 \ldots \mu_{s}\lambda}{}^{\lambda}=0$. 
This last constraint only makes sense for $s\geq 3$. Our conventions regarding symmetrisation {\it etc.}\ 
are explained in appendix~\ref{app:A}.

Now we will count the number of independent components of the fields. Recall that a completely symmetric 
tensor of rank $s$ in three dimensions has ${(s+1)(s+2)\over 2}$ independent components. In our case, 
because of the double trace constraint, many of these components are not actually independent. 
The constraints are as many in number as those of a symmetric tensor of rank $(s-4)$. Therefore 
the net number of independent components is given by 
\begin{equation}
{(s+1)(s+2)\over 2}-{(s-3)(s-2)\over 2}=4s-2\ .
\end{equation}

We now argue that half of these are gauge degrees of freedom. Recall that the gauge parameter
is given by a traceless rank $(s-1)$ symmetric tensor $\xi_{\mu_1\mu_2\ldots \mu_{s-1}}$. The number 
of independent components of $\xi_{(s-1)}$ is therefore
\begin{equation}
{s(s+1)\over 2}-{(s-1)(s-2)\over 2}=2s-1\ .
\end{equation}
Therefore the non-gauge components are also $(2s-1)$ in number.
This fits in with what we expect for these fields, namely that they have no 
net propagating degrees of freedom (at least in the bulk of AdS$_3$).

Let us now analyse the representation theoretic content of these different modes. 
This will give us an important clue to the analysis of the one loop answer. 
We can decompose the field $\vp_{(s)}$ in the following way
\be{phidecomp}
\vp_{\mu_1\mu_2 \ldots \mu_s}=
\vp^{\rm TT}_{\mu_1\mu_2 \ldots \mu_s}+g_{(\mu_1\mu_2}\tilde{\vp}_{\mu_3\ldots \mu_s)}
+\nabla_{(\mu_1}\xi_{\mu_2 \ldots \mu_{s})} \ .
\ee
Here $\vp^{\rm TT}_{(\mu_1\mu_2 \ldots \mu_s)}$ is the transverse, traceless piece of $\vp_{(s)}$
and consists of two independent components carrying helicity $\pm s$. 
$\tilde{\vp}_{(\mu_3\ldots \mu_s)}$ is the spin $(s-2)$ piece  which carries all the trace information 
of $\vp_{(s)}$. Finally, $\xi_{(s-1)}$ are the gauge parameters. Note that the double
trace constraint on $\varphi_{(s)}$ of eq.\ (\ref{dbtr}) implies that $\tilde\varphi_{(s-2)}$ is traceless. 

In what follows it will be important for us to make the further decomposition of the gauge
field $\xi_{(s-1)}$ into its traceless transverse component $\xi^{\rm TT}$, as well as
\begin{equation}\label{xidecomp}
\xi_{\mu_1\ldots\mu_{s-1}} = 
\xi^{\rm TT}_{\mu_1\ldots\mu_{s-1}}  + \xi^{(\sigma)}_{\mu_1\ldots \mu_{s-1}} \ , 
\end{equation}
where $\xi^{(\sigma)}_{(s-1)}$ is the longitudinal component, that can be written as 
\begin{equation}\label{xis}
\xi^{(\sigma)}_{\mu_1\ldots \mu_{s-1}}  = 
\nabla_{(\mu_1} \sigma_{\mu_2\ldots\mu_{s-1})} - \frac{2}{(2s-3)}\,
g_{(\mu_1\mu_2} \nabla^{\lambda}
\sigma_{\mu_3\ldots\mu_{s-1})\lambda} \ , 
\end{equation}
with $\sigma_{(s-2)}$ a traceless symmetric tensor. The transverse, traceless 
component $\xi^{\rm TT}_{(s-1)}$ carries helicity $\pm (s-1)$.

In order to exhibit the remaining helicitiy components we can now further decompose 
$\tilde\vp_{(s-2)}$ and $\sigma_{(s-2)}$ into their transverse traceless, as well as their
longitudinal 
spin $(s-3)$ components. 
%
The longitudinal pieces
that appear in either of these decompositions can then again be decomposed into transverse 
traceless spin $(s-3)$ components, 
together with longitudinal components of spin $(s-4)$, {\it etc}. In this way we can see that both 
$\tilde\vp_{(s-2)}$ and $\sigma_{(s-2)}$ have
helicity components corresponding to all the helicities less or equal to $(s-2)$; this gives
rise to $2(s-2)+1=2s-3$ components for each of the two fields.  In summary we therefore have 
\begin{list}{(\roman{enumi})}{\usecounter{enumi}}
\item a symmetric transverse traceless field  $\varphi^{\rm TT}_{(s)}$ of spin $s$,
with helicities $\pm s$ [2 components]
\item a symmetric transverse traceless gauge mode $\xi^{\rm TT}_{(s-1)}$ of spin $s-1$, 
with helicities $\pm (s-1)$ [2 components]
\item a symmetric traceless (but not necessarily transverse) field $\tilde\varphi_{(s-2)}$ of spin $s-2$,
with helicities $0,\pm 1, \pm 2, \ldots, \pm (s-2)$ [$2s-3$ components]
\item a symmetric traceless (but not necessarily transverse) gauge field  $\sigma_{(s-2)}$ 
of spin $s-2$, with helicities $0,\pm 1, \pm 2, \ldots, \pm (s-2)$ [$2s-3$ components]
\end{list}
In particular, there are therefore $2s-1$ non-gauge and $2s-1$ gauge components, in agreement
with the above counting. Note that there are precisely as many gauge components in 
$\sigma_{(s-2)}$, as there are components in $\tilde\vp_{(s-2)}$. In fact, if we 
consider the trace part of  (\ref{phidecomp}), the tracelessness of $\sigma_{(s-2)}$ implies that 
\begin{eqnarray}
\vp_{\mu_1\mu_2 \ldots \mu_{s-2} \lambda}{}^{\lambda}  &= &(2s-1)\, \tilde\varphi_{\mu_1\ldots \mu_{s-2}} 
+\nabla^{\lambda}\xi^{(\sigma)}_{\mu_1\ldots \mu_{s-2}\lambda} \cr & =&
(2s-1)\, \tilde\varphi_{\mu_1\ldots \mu_{s-2}} + ({\cal K} \sigma)_{\mu_1\ldots \mu_{s-2}}  \ ,
\end{eqnarray}
where ${\cal K}$ is a linear second order differential operator.
Thus, at least classically, we can gauge away $\tilde\vp_{(s-2)}$ completely \cite{Mikhailov:2002bp}. 
This therefore suggests
that in the calculation of the one loop determinant, the $\tilde\vp_{(s-2)}$ and $\sigma_{(s-2)}$
fields will give cancelling contributions. The final answer should therefore only involve
the helicity $\pm s$ non-gauge modes of $\vp^{\rm TT}_{(s)}$,  as well as the helicity $\pm (s-1)$ 
gauge modes of $\xi^{\rm TT}_{(s-1)}$. This is the intuitive explanation of the answer \eq{dets}.
Below we will see explicitly how this happens from a careful consideration of the quadratic functional 
integral for the $\vp$-field. 
 
\section{The general setup}

The quadratic fluctuations we are interested in can be computed from the functional integral 
\be{phisint}
Z^{(s)}={1\over {\rm Vol(gauge}\, {\rm group)}}\, \int [D\vp_{(s)}] \, e^{-S[\vp_{(s)}]} \ .
\ee
Here $S[\vp_{(s)}]$ is the action of a spin-$s$ field in a $D=3$ dimensional AdS 
background \cite{Fronsdal:1978rb} --- we are using the conventions and notations of 
\cite{Campoleoni:2010zq}
\begin{equation}\label{action1}
S[\vp_{(s)}] = \int d^3x \sqrt{g}\, \vp^{\mu_1\ldots \mu_s} \left( 
\widehat{\cal F}_{\mu_1\ldots\mu_s} - \frac{1}{2} g_{(\mu_1\mu_2} 
\widehat{\cal F}_{\mu_3\ldots \mu_s)\lambda}{}^{\lambda} \right) \ ,
\end{equation}
where the symmetric tensor $\widehat{\cal F}_{\mu_1\ldots\mu_s}$ is defined in $D=3$ as 
\begin{equation}
\widehat{\cal F}_{\mu_1\ldots\mu_s}= 
{\cal F}_{\mu_1\ldots\mu_s} - \frac{s(s-3)}{\ell^2} \, \vp_{\mu_1\ldots\mu_s} - 
\frac{2}{\ell^2} \,g_{(\mu_1 \mu_2}  \vp_{\mu_3\ldots\mu_s)\lambda}{}^{\lambda} \ .
\end{equation}
Finally, ${\cal F}_{\mu_1\ldots\mu_s}$ equals\footnote{Note that in the last term there is a factor of
$\frac{1}{2}$ relative to eq.\ (2.1) of \cite{Campoleoni:2010zq}. This arises because covariant derivatives
do not in general commute, and thus we need to sum over both orderings of $\mu_1,\mu_2$. Our
conventions regarding symmetrisation are explained in appendix~\ref{app:A}.}
\begin{equation}\label{vpf1}
{\cal F}_{\mu_1\ldots\mu_s} = 
\Delta \vp_{\mu_1\ldots\mu_s} - 
\nabla_{(\mu_1} \nabla^{\lambda} \vp_{\mu_2\ldots\mu_s)\lambda} + 
\frac{1}{2} 
\nabla_{(\mu_1} \nabla_{\mu_2} \vp_{\mu_3\ldots\mu_s)\lambda}{}^{\lambda} \ .
\end{equation}
We should mention in passing that the case $s=2$ reduces to the familiar gravity action on
AdS$_3$, see for example \cite{Grumiller:2010xv}.

In order to evaluate the path integral $Z^{(s)}$ in (\ref{phisint})  it is useful to change variables as 
\be{ghost}
[D\vp_{(s)}]=Z_{\rm gh}^{(s)}\, [D\vp^{\rm TT}_{(s)}]\, [D\tilde{\vp}_{(s-2)}]\, [D\xi_{(s-1)}] \ ,
\ee
where we use the same decomposition as in (\ref{phidecomp}). Here $Z_{\rm gh}^{(s)}$ denotes
the ghost determinant that arises from the change of variables. 

The gauge invariance of the action, together with the orthogonality of the first terms of (\ref{phidecomp}), 
implies that 
\be{sdecomp}
S[\vp_{(s)}]=S[\vp^{\rm TT}_{(s)}]+S[\tilde{\vp}_{(s-2)}] \ ,
\ee
and the first term is simply
\be{phitt}
S[\vp^{\rm TT}_{(s)}] = \int d^3x\sqrt{g}\, \vp^{\rm TT\; \mu_1\ldots \mu_s}\, 
\left(-\Delta+\frac{s(s-3)}{\ell^2}\right)\, \vp^{\rm TT}_{\mu_1\ldots \mu_s}\ .
\ee
Thus the functional integral over the TT modes is easily evaluated to be 
\be{funcsimp}
Z^{(s)}=Z_{\rm gh}^{(s)}\, \left[\det\left(-\Delta+\frac{s(s-3)}{\ell^2}\right)_{(s)}^{\rm TT}\right]^{-\half}\, 
\int [D\tilde{\vp}_{(s-2)}]\, e^{-S[\tilde{\vp}_{(s-2)}]}\ .
\ee
The determination of the functional integral requires therefore that we compute 
$Z_{\rm gh}^{(s)}$, as well as the quadratic integral over $\tilde{\vp}_{(s-2)}$. 
Let us briefly discuss both terms. 

\subsection{The quadratic action of $\tilde\vp_{(s-2)}$}\label{sec:2.1}

For the component of $\varphi_{(s)}$ proportional to $\tilde\vp_{(s-2)}$, see eq.\ 
(\ref{phidecomp}), the above action simplifies considerably. Indeed, it follows 
directly from (\ref{action1}) that 
\begin{equation}\label{vpt1}
S[\tilde{\vp}_{(s-2)}] = - \frac{s(s-1)(2s-3)}{4}\, 
\int d^3x \sqrt{g}\, 
\tilde\vp^{\mu_1\ldots\mu_{s-2}} \, \widehat{\cal F}_{\mu_1\ldots \mu_{s-2}\lambda}{}^{\lambda} \ ,
\end{equation}
where $\widehat{\cal F}$ is evaluated on $\vp_{(s)}=g\, \tilde\vp_{(s-2)}$. By an explicit computation
one finds that 
\bea{hatf}
\hat{\cal F}_{\mu_1\ldots \mu_s}(\vp)&=&g_{(\mu_1\mu_2}\Biggl[\Delta\, \vptl_{\, \mu_3\ldots \mu_s)}
-\nabla_{\mu_3}\nabla^{\lambda}\vptl_{\mu_4\ldots \mu_s)\lambda} 
-\frac{(s^2+s-2)}{\ell^2}\vptl_{\, \mu_3\ldots \mu_s)} \Biggr]  \cr
& & \qquad \qquad +{(2s-3)\over 2}\nabla_{(\mu_1}\nabla_{\mu_2}\vptl_{\, \mu_3\ldots \mu_s)} \ .
\eea
Therefore 
$\hat{\cal F}_{\mu_3\ldots \mu_s\lambda}{}^{\lambda} =g^{\mu_1\mu_2}\hat{\cal F}_{\mu_1\ldots \mu_s}$ 
is given by
\begin{equation}\label{trhatf}
\begin{array}{rcl}
\hat{\cal F}_{\mu_3\ldots \mu_s\lambda}{}^{\lambda} &=& 
(2s-1)\Biggl[\Delta\vptl_{( \mu_3\ldots \mu_s)}
-\nabla_{(\mu_3}\nabla^{\lambda}\vptl_{\mu_4\ldots \mu_s)\lambda} 
-\frac{(s^2+s-2)}{\ell^2}\vptl_{(\mu_3\ldots \mu_s)}\Biggr] \\
&&\qquad \qquad -2g_{(\mu_3\mu_4}\nabla^{\lambda}\nabla^{\nu}
\tilde\vp_{\mu_5\ldots \mu_s)\lambda\nu} 
+{(2s-3)\over 2}g^{\lambda\nu}\nabla_{(\lambda}\nabla_{\nu}\vptl_{\, \mu_3\ldots \mu_s)} \ .
\end{array}
\end{equation}
Note that the second but last term in \eq{trhatf} will not contribute when put into \eq{vpt1} because of 
the tracelessness condition on $\vptl$. The last term in \eq{trhatf} can be evaluated to be
\begin{equation}\label{lasthatf}
\begin{array}{rcl}
g^{\mu_1\mu_2}\nabla_{(\mu_1}\nabla_{\mu_2}\vptl_{\, \mu_3\ldots \mu_s)} &=&
 2\Delta\vptl_{(\mu_3\ldots \mu_s)} 
+ 2\nabla_{(\mu_3}\nabla^{\lambda}\vptl_{\mu_4\ldots \mu_s)\lambda}
+2\nabla^{\lambda}\nabla_{(\mu_3}\vptl_{\mu_4\ldots \mu_s)\lambda}\\[4pt]
&=&  2\Delta\vptl_{(\mu_3\ldots \mu_s)} 
+ 4\nabla_{(\mu_3}\nabla^{\lambda}\vptl_{\mu_4\ldots \mu_s)}
-\frac{(s-1)(s-2)}{\ell^2}\vptl_{\, \mu_3\ldots \mu_s)}\ . 
\end{array}
\end{equation}
Plugging (\ref{lasthatf}) and (\ref{trhatf}) into \eq{hatf},  the quadratic action for $\vptl$ finally becomes
\bea{quadvp1}
S[\tilde\vp_{(s-2)}] & = & C_s\, 
\int d^3x\sqrt{g}\, \tilde\vp^{\, \mu_3\ldots \mu_s}\, \hat{\cal F}_{\mu_3\ldots \mu_s\lambda}{}^{\lambda} \cr
&= & C_s \int d^3x\sqrt{g} \,\tilde\vp^{\,\mu_3\ldots \mu_s} 
\Biggl[ 4(s-1)\left(\Delta -\frac{s^2-s+1}{\ell^2}\right)\tilde\vp_{\mu_3\ldots \mu_s} \\
& & \quad \qquad \qquad\qquad \qquad+(2s-5)\nabla_{(\mu_3}
\nabla^{\lambda}\tilde\vp_{\mu_4\ldots \mu_s)\lambda}  \Biggr]  \ , \nonumber
\eea
where $C_s$ is an (unimportant) constant. The path integral over $\tilde\vp_{(s-2)}$ is now
straightforward, and can be expressed in terms of the determinant of the differential operator 
appearing in (\ref{quadvp1}).  Notice, however, that $\tilde\vp_{(s-2)}$ is only traceless,
and not transverse. If we want to express this determinant in terms of differential operators
acting on transverse traceless operators, more work will be required. This will be sketched for
$s=3$ below in section~\ref{sec:3.2}.

\subsection{The ghost determinant}

For the evaluation of the ghost determinant we shall follow the same strategy as in 
\cite{Grumiller:2010xv}. This is to say, we write
\begin{eqnarray}\label{Z1}
1 & = & \int [D\varphi_{(s)}]\, e^{-\langle \varphi_{(s)},\varphi_{(s)} \rangle} \nonumber \\
& = & Z_{\rm gh}^{(s)} 
\int [D\vp^{\rm TT}_{(s)}]\, [D\tilde{\vp}_{(s-2)}]\, [D\xi_{(s-1)}]\,
e^{-\langle \varphi_{(s)},\varphi_{(s)} \rangle} \ ,
\end{eqnarray}
where $\varphi_{(s)}\equiv \varphi_{(s)}(\vp^{\rm TT}_{(s)},\tilde{\vp}_{(s-2)},\xi_{(s-1)})$
as in (\ref{phidecomp}). Next we expand out 
\begin{eqnarray}
\langle \varphi_{(s)},\varphi_{(s)} \rangle & = & 
\langle \varphi^{\rm TT}_{(s)},\varphi^{\rm TT}_{(s)} \rangle + 
\langle g\tilde{\vp}_{(s-2)}, g\tilde{\vp}_{(s-2)}\rangle  +
\langle \nabla \xi_{(s-1)}, \nabla \xi_{(s-1)} \rangle \nonumber \\
& & \quad + 
\langle  g\tilde{\vp}_{(s-2)}, \nabla \xi_{(s-1)} \rangle  + 
\langle  \nabla \xi_{(s-1)} ,  g\tilde{\vp}_{(s-2)} \rangle
\ .
\end{eqnarray}
In order to remove the off-diagonal terms of the last line, we rewrite (\ref{phidecomp}) as 
\begin{equation}\label{phidecomp1}
\vp_{\mu_1\mu_2 \ldots \mu_s}=
\vp^{\rm TT}_{\mu_1\mu_2 \ldots \mu_s}+g_{(\mu_1\mu_2}\tilde{\vp}'_{\mu_3\ldots \mu_s)}
+\left(\nabla_{(\mu_1}\xi_{\mu_2 \ldots \mu_{s})}  - {\textstyle \frac{2}{2s-1}}
g_{(\mu_1\mu_2} \, \nabla^\lambda \xi_{\mu_3\ldots \mu_{s}) \lambda} \right) \ ,
\end{equation}
where
\begin{equation}
\tilde{\vp}'_{\mu_1\ldots \mu_{s-2}} = \tilde{\vp}_{\mu_1\ldots \mu_{s-2}} + 
{\textstyle \frac{2}{2s-1}} \nabla^\lambda \xi_{\mu_1\ldots \mu_{s-2} \lambda} \ .
\end{equation}
Then the quadratic term takes the form
\begin{eqnarray}
\langle \varphi_{(s)},\varphi_{(s)} \rangle  & =  & 
\langle \varphi^{\rm TT}_{(s)},\varphi^{\rm TT}_{(s)} \rangle + 
\langle g\tilde{\vp}'_{(s-2)}, g\tilde{\vp}'_{(s-2)}\rangle \nonumber \\
& &  +
\langle (\nabla \xi_{(s-1)} -{\textstyle \frac{2}{2s-1}} g \nabla \xi), 
(\nabla \xi_{(s-1)} - {\textstyle\frac{2}{2s-1}} g \nabla \xi) \rangle 
\ . 
\end{eqnarray}
Both the $\varphi^{\rm TT}_{(s)}$ and the $\tilde{\vp}'_{(s-2)}$ path integral are now
trivial, as is the Jacobian coming from the change of measure in going from $\tilde{\vp}_{(s-2)}$ 
to $\tilde{\vp}'_{(s-2)}$. Thus the ghost determinant simply becomes 
\begin{equation}\label{Z2}
\frac{1}{Z_{\rm gh}^{(s)}} = 
\int [D\xi_{(s-1)}]\, e^{-\langle (\nabla \xi_{(s-1)} -{\textstyle \frac{2}{2s-1}} g \nabla \xi), 
(\nabla \xi_{(s-1)} - {\textstyle\frac{2}{2s-1}} g \nabla \xi) \rangle } \ .
\end{equation}
The exponent can be simplified further by integrating by parts to get 
\begin{eqnarray}\label{gh1}
S_\xi & = & \langle (\nabla \xi_{(s-1)} -{\textstyle \frac{2}{2s-1}} g \nabla \xi), 
(\nabla \xi_{(s-1)} - {\textstyle\frac{2}{2s-1}} g \nabla \xi) \rangle   \nonumber \\
& = & s \int d^3z \sqrt{g} \Bigl[ 
\xi_{\mu_1\ldots \mu_{s-1}} \left( - \Delta + \frac{s(s-1)}{\ell^2} \right) \xi^{\mu_1\ldots \mu_{s-1}} \nonumber \\
& & \qquad \qquad \qquad
- \frac{(s-1) (2s-3)}{(2s-1)}\,  \xi_{\mu_1\ldots \mu_{s-2}\lambda} \nabla^{\lambda} \nabla_{\nu} 
\xi^{\mu_1\ldots \mu_{s-2}\nu} \Bigr]\ .
\end{eqnarray}
This is the path integral we have to perform. Before doing the calculation in the general case, it is 
instructive to analyse the simplest case, $s=3$, first. The impatient reader is welcome to skip the 
next section and proceed directly to Sec.~4 where we perform the general analysis of the ghost 
determinant.

\section{The Case of Spin Three}\label{sec:3}

As explained in the previous section, see eq.\ (\ref{funcsimp}), the calculation of the 1-loop determinant
is reduced to determining the ghost determinant (\ref{Z2}) as well as the determinant arising from 
(\ref{quadvp1}). We shall first deal with the ghost determinant.

\subsection{Calculation of the Ghost Determinant}

For $s=3$ the $\xi$-dependent exponent of  (\ref{gh1}) is of the form
\begin{equation}\label{quad1}
S_\xi = 3 \int d^3x \sqrt{g}  \, \Bigl[ 
\xi_{\nu\rho} \left(- \Delta + \frac{6}{\ell^2} \right) \xi^{\nu\rho} 
- {6\over 5}\, \xi_\nu{}^{\rho} \nabla_\rho \nabla_\mu \xi^{\mu\nu} \Bigr] \ .
\end{equation}
We shall first do the calculation in a pedestrian manner, following the same methods as in
\cite{Grumiller:2010xv}. We shall then explain how our result can be more efficiently obtained. 
To start with we decompose $\xi^{\mu\nu}$ as 
\begin{equation}\label{decomp1}
\xi^{\mu\nu} = \xi^{{\rm TT}\, \mu\nu} + \nabla^\mu \sigma^{{\rm T}\nu} + \nabla^\nu \sigma^{{\rm T}\mu} 
+ \left( \nabla^{\mu} \nabla^{\nu} - \frac{1}{3} g^{\mu\nu} \nabla^2 \right) \psi 
= \xi^{{\rm TT}\, \mu\nu} + \nabla^{(\mu} \sigma^{{\rm T}\nu)} + \psi^{\mu\nu} \ , 
\end{equation}
where $ \xi^{{\rm TT}\, \mu\nu}$ is the transverse and traceless part of $\xi^{\mu\nu}$,
while $\sigma^{{\rm T}\nu}$ is (traceless) 
and transverse, {\it i.e.}\ 
\begin{equation}\label{strans}
\nabla_\nu \sigma^{{\rm T}\nu} = 0 \  .
\end{equation}
Plugging (\ref{decomp1}) into (\ref{quad1}) we obtain after a lengthy calculation --- some of the 
details are explained in appendix~\ref{app:3de} ---
\begin{eqnarray}
S_\xi & = &  \int d^3x \sqrt{g}  \, \Biggl[ 
3\, \xi^{{\rm TT}}_{\nu\rho}\,  \left(- \Delta + \frac{6}{\ell^2} \right) \xi^{{\rm TT}\, \nu\rho} \nonumber \\
& & \qquad  \qquad \quad 
+ {48\over 5} \, \sigma^{\rm T}_{\nu} \left(-\Delta + \frac{2}{\ell^2} \right) \, 
\left(-\Delta + \frac{7}{\ell^2} \right)  \, \sigma^{{\rm T}\nu} \nonumber \\
& & \qquad \qquad \quad  
+ \frac{18}{5}\, \psi \left( - \Delta \right) \,  
\left(-\Delta + \frac{3}{\ell^2} \right) \, \left(-\Delta + \frac{8}{\ell^2} \right)
\, \psi \Biggr] 
\ .\label{Sxi}
\end{eqnarray}
The ghost determinant  (\ref{Z2}) is then simply 
\begin{eqnarray}\label{Z3}
Z_{\rm gh}^{(s=3)} & =  & 
J_1^{-1} \, \Biggl\{\det \left( - \Delta + \frac{6}{\ell^2} \right)^{\rm TT}_{(2)} \, 
\det \left[\left(-\Delta + \frac{2}{\ell^2} \right) \,  \left(-\Delta + \frac{7}{\ell^2} \right)\right]^{\rm T}_{(1)} \nonumber \\
& & \qquad\,\, \times
\det \left[  \left( - \Delta \right) \,  
\left(-\Delta + \frac{3}{\ell^2} \right) \, \left(-\Delta + \frac{8}{\ell^2}\right) \right]_{(0)} \Biggr\}^{\frac{1}{2}}\ , 
\end{eqnarray}
where $J_1$ is the Jacobian from the change of measure in going from
$\xi$ to $(\xi^{\rm TT},\sigma^{\rm T},\psi)$. This can be calculated from the identity
\begin{equation}
1 = \int D\xi e^{-\langle \xi, \xi \rangle} = 
      \int J_1 D\xi^{\rm TT} D\sigma^{\rm T} d\psi \,
      e^{-\langle \xi(\xi^{\rm TT},\sigma^{\rm T},\psi),\, \xi(\xi^{\rm TT},\sigma^{\rm T},\psi) \rangle} \ . 
\end{equation}
Expanding out the exponential, the terms of interest are
\begin{equation}\label{sigma1}
\int d^3x \sqrt{g} \,\,  \nabla_{(\mu} \sigma^{\rm T}_{\nu)} \, \nabla^{(\mu} \sigma^{{\rm T}(\nu)} = 
- 2 \int \sigma^{\rm T}_\nu \left( \Delta - \frac{2}{\ell^2} \right) \sigma^{{\rm T}\nu} \ .
\end{equation}
and 
\begin{eqnarray}
& & \int d^3x \sqrt{g} \,\, 
\Biggl[\left( \nabla_\mu \nabla_\nu - \frac{1}{3} g_{\mu\nu} \Delta \right) \psi \Biggr]\, 
\Biggl[\left( \nabla^\mu \nabla^\nu - \frac{1}{3} g^{\mu\nu} \Delta \right)  \psi \Biggr] \nonumber \\
&& \qquad \qquad 
=  \frac{2}{3} \,  \int d^3x \sqrt{g} \,\, \psi \,\, \left[ \left( -\Delta \right)\, \left( - \Delta + \frac{3}{\ell^2} \right) \right]
\psi \ . 
\end{eqnarray}
Thus the $(-\Delta + \frac{2}{\ell^2} )$ term is cancelled from the first line of (\ref{Z3}) and similarly the 
$(-\Delta) (-\Delta + \frac{3}{\ell^2} )$ term from the second line. The complete ghost determinant 
for $s=3$ therefore equals
\begin{equation}\label{ghost1}
Z_{\rm gh}^{(s=3)} = \Biggl[ \det\left(-\Delta +\frac{6}{\ell^2} \right)^{\rm TT}_{(2)} \, 
\det\left(-\Delta + \frac{7}{\ell^2} \right)^{\rm T}_{(1)} \,   \left(-\Delta + \frac{8}{\ell^2} \right)_{(0)} \,
\Biggr]^{\frac{1}{2}} \ . 
\end{equation}
\medskip

For the following it will be important to observe that this result can be obtained 
more directly. Indeed, as the above calculation has demonstrated, there are many
cancellations between terms arising from $S_{\xi}$ and the change of measure $J_1$. 
Actually, it is not difficult to see how this comes about. Consider for example the vector 
part. The factor by which the second line of (\ref{Sxi}) differs from (\ref{sigma1}) is the eigenvalue 
of the differential operator
\begin{equation}\label{operator1}
\bigl( {\cal L}^{(3)} \xi \bigr)^{\nu\rho} \equiv 
\left(-\Delta + \frac{6}{\ell^2} \right) \xi^{\nu\rho}  - \frac{3}{5} 
\left(\nabla^{\rho} \nabla_{\mu} \xi^{\nu\mu} + \nabla^{\nu} \nabla_{\mu} \xi^{\rho\mu}  \right)
\end{equation}
evaluated on the tensors of the form $\xi^{\nu\rho} = \nabla^{(\nu} \sigma^{{\rm T}\rho)}$. This
reproduces indeed (\ref{ghost1}) since
we find 
\begin{eqnarray}
& & \left(-\Delta + \frac{6}{\ell^2} \right) \nabla^{(\nu} \sigma^{{\rm T}\rho)} 
- \frac{3}{5} 
\left( \nabla^\rho \nabla_\mu \nabla^{(\mu}\sigma^{{\rm T}\nu)} 
+ \nabla^\nu \nabla_\mu \nabla^{(\mu}\sigma^{{\rm T}\rho)} \right)
\nonumber \\
& & \qquad \qquad
= \frac{8}{5} \Bigl[ \nabla^\nu \left( - \Delta + \frac{7}{\ell^2} \right) \sigma^{{\rm T}\rho} + 
\nabla^\rho \left( - \Delta + \frac{7}{\ell^2} \right) \sigma^{{\rm T}\nu} \Bigr] \ . \label{eigen1}
\end{eqnarray}
Similarly, the action of ${\cal L}^{(3)}$ on $\psi^{\mu\nu}$  
leads to 
\begin{eqnarray}
& & \left(-\Delta + \frac{6}{\ell^2} \right) \psi_{\nu\rho} 
- \frac{6}{10} \left(
\nabla_\rho \nabla^\mu \psi_{\mu\nu} + \nabla_\nu \nabla^\mu \psi_{\mu\rho}\right) \nonumber \\
& & \qquad \qquad 
= \frac{9}{5} \nabla_\rho \nabla_\nu \left( - \Delta + \frac{8}{\ell^2} \right)
\psi - \frac{1}{3} g_{\rho\nu} \, \Delta \left (-\Delta + \frac{12}{\ell^2} \right) \psi  \ .
\end{eqnarray}
While $\psi^{\mu\nu}$ is not an eigenvector of (\ref{operator1}), the second term proportional to 
$g_{\rho\nu}$ does not actually matter for the 1-loop calculation since the result is contracted with
the traceless tensor $\psi_{\rho_\nu}$, see eq.\ (\ref{quad2}).

\subsection{The Quadratic Contribution from $\tilde\vp$}\label{sec:3.2}

The other piece of the calculation is the quadratic contribution from $\tilde\vp$, which for $s=3$
takes the form
\begin{equation}\label{comp}
S[\tilde\vp_{(1)}] =  \frac{9}{2} \int d^3 \sqrt{g} \,  
\Biggl[ 8\, \tilde\vp^\rho \left(-\Delta + \frac{7}{\ell^2} \right) \tilde\vp_\rho 
-\tilde\vp^\rho \, \nabla_\rho \nabla^\lambda \tilde\vp_{\lambda} \Biggr] \ .
\end{equation}
In order to express the determinant in terms of those acting on traceless transverse components, 
we now decompose  $\tilde\vp_\rho$  into its transverse and longitudinal component 
\begin{equation}
\tilde\vp_\rho = \tilde\vp^{\rm T}_\rho + \nabla_\rho \chi \ .
\end{equation}
By the usual argument the above quadratic action then becomes
\begin{eqnarray}
S& =  & - \frac{9}{2} \int d^3 \sqrt{g} \,  
\Biggl[ 8\, \tilde\vp^{{\rm T}\, \rho} \left(\Delta - \frac{7}{\ell^2} \right) \tilde\vp^{\rm T}_\rho \nonumber \\
& & \qquad \qquad \qquad 
- 8\, \chi \nabla^\rho \left(\Delta - \frac{7}{\ell^2} \right) \nabla_\rho \chi 
- \chi\, \nabla^\rho \, \nabla_\rho \nabla^\lambda \nabla_\lambda \chi \Biggr] \nonumber \\
& = &   \frac{9}{2} \int d^3 \sqrt{g} \,  
\Biggl[ 8\, \tilde\vp^{{\rm T}\, \rho} \left(-\Delta + \frac{7}{\ell^2} \right) \tilde\vp^{\rm T}_\rho
+ 9 \chi \left( - \Delta + \frac{8}{\ell^2} \right) (- \Delta)  \chi \Biggr]\ . 
\end{eqnarray}
The last $(-\Delta)$ factor is removed by the Jacobian that arises because of the change of 
variables $\tilde\vp \equiv \tilde\vp (\tilde\vp^{\rm T},\chi)$; the relevant term there is simply 
\begin{equation}
\int d^3x \sqrt{g} \, (\nabla^\rho \chi ) (\nabla_\rho \chi) =  \int d^3x \sqrt{g} \,  \chi  (-\Delta \chi)  \ .
\end{equation}
The correction term coming from this part of the calculation is therefore of the form
\begin{equation}
Z_{\tilde\vp_{(1)}}^{(s=3)} = \Biggl[ \det  \left(-\Delta + \frac{7}{\ell^2} \right)^{T}_{(1)} \, 
\det \left( - \Delta + \frac{8}{\ell^2} \right)_{(0)} \Biggr]^{-\frac{1}{2}} \ .
\end{equation}
This cancels precisely against two of the factors in $Z_{\rm gh}^{(s=3)}$ of eq.\
(\ref{ghost1}), as expected from our general considerations above. 
Combining the different pieces as in eq.\ (\ref{funcsimp}), the 
total 1-loop determinant for $s=3$ then equals
\begin{equation}
Z^{(s=3)} =  \Biggl[ \det\left(-\Delta  \right)^{\rm TT}_{(3)}\Biggr]^{-\frac{1}{2}}  \, 
\Biggl[ \det\left(-\Delta + \frac{6}{\ell^2} \right)^{\rm TT}_{(2)}\Biggr]^{\frac{1}{2}}  \ .
\end{equation}
As expected, only the helicity $s$ and helicity $(s-1)$ terms therefore 
contribute to this determinant.

\section{Quadratic Fluctuations for General Spin}

The above calculation is fairly technical, and we cannot hope to generalise it directly to
higher spin. However, as explained above, we expect that the contributions of $\tilde\vp_{(s-2)}$ 
and $\sigma_{(s-2)}$ should cancel each other, and it should therefore be possible
to organise the calculation in a way in which this becomes manifest. In the following
we shall explain how this can be achieved. In particular, we shall explain that most
of the ghost determinant will actually just cancel the quadratic contribution from 
$\tilde\vp_{(s-2)}$.

\subsection{The ghost determinant}

Generalising the definition of ${\cal L}^{(3)}$ in (\ref{operator1}) let us define ${\cal L}^{(s)}$
to be the differential operator ${\cal L}^{(s)}$ appearing in the integral (\ref{gh1})
\begin{equation}\label{Ldef}
\bigl({\cal L}^{(s)} \xi \bigr)_{\mu_1\ldots \mu_{s-1}} =
\left( - \Delta + \frac{s(s-1)}{\ell^2} \right) \xi_{\mu_1\ldots \mu_{s-1}} 
- \frac{(2s-3)}{(2s-1)} \nabla_{(\mu_1} \nabla^{\lambda} \xi_{\mu_2\ldots\mu_{s-1})\lambda} \ .
\end{equation}
Let us separate $\xi$ into its transverse traceless component as well as $\sigma_{(s-2)}$ as
in \eq{xidecomp}, {\it i.e.}\ $\xi_{(s-1)} = \xi^{\rm TT}_{(s-1)} +  \xi^{(\sigma)}_{(s-1)}$ with 
\begin{equation}\label{xis1}
\xi^{(\sigma)}_{\mu_1\ldots \mu_{s-1}}  = 
\nabla_{(\mu_1} \sigma_{\mu_2\ldots\mu_{s-1})} - \frac{2}{(2s-3)}\,
g_{(\mu_1\mu_2} \nabla^{\lambda}
\sigma_{\mu_3\ldots\mu_{s-1})\lambda} \ .
\end{equation}
On the transverse traceless components $\xi^{\rm TT}_{(s-1)}$ of $\xi_{(s-1)}$ the second term 
of ${\cal L}^{(s)}$ vanishes, and the operator has a simple form, namely
\begin{equation}\label{xitrans}
{\cal L}^{(s)} \xi^{\rm TT}_{(s-1)}  = \left(-\Delta + \frac{s(s-1)}{\ell^2} \right) \xi^{\rm TT}_{(s-1)} \ .
\end{equation}
In order to determine ${\cal L}^{(s)}  \xi_{(s-1)}$ it  therfore remains to calculate 
${\cal L}^{(s)}  \xi^{(\sigma)}_{(s-1)}$ as a differential operator on 
$\sigma_{(s-2)}$. Since the resulting expression will be contracted with $\xi^{(\sigma)}_{(s-1)}$, we 
only have to evaluate the operator up to `trace terms', ({\it i.e.}\ terms that are proportional to
$g_{\mu_i \mu_j}$ for some indices $i,j\in \{1,\ldots,s-1\}$). The calculation can be broken up into
different terms. From the first term of ${\cal L}^{(s)}$ we get 
\begin{equation}\label{a1}
\left(-\Delta + \frac{s(s-1)}{\ell^2} \right) \nabla_{(\mu_1} \sigma_{\mu_2\ldots \mu_{s-1})} 
\cong \nabla_{(\mu_1} \left( -\Delta + \frac{(s-1)(s+2)}{\ell^2} \right) \, \sigma_{\mu_2\ldots \mu_{s-1})} \ ,
\end{equation}
where $\cong$ always denotes equality up to trace terms. The action of $(-\Delta + \frac{s(s-1)}{\ell^2})$
on the second term in (\ref{xis}) only produces a trace term. 

This leaves us with evaluating the second term of ${\cal L}^{(s)}$. The relevant formulae are 
\begin{eqnarray}
- \frac{(2s-3)}{(2s-1)}  \nabla_{(\mu_1} \nabla^{\lambda} 
\nabla_{(\mu_2} \sigma_{\mu_3\ldots\mu_{s-1}\lambda)} & \cong &
\frac{(2s-3)}{(2s-1)} 
\nabla_{(\mu_1} \left( -\Delta + \frac{(s-1)(s-2)}{\ell^2} \right) \, \sigma_{\mu_2\ldots \mu_{s-1})}  \nonumber \\
& & \quad - 
\frac{(2s-3)}{(2s-1)}  
\nabla_{(\mu_1} \nabla_{\mu_2} \nabla^\lambda \sigma_{\mu_3\ldots\mu_{s-1})\lambda} \ ,\label{a2}
\end{eqnarray}
as well as 
\begin{equation}\label{a3}
- \frac{(2s-3)}{(2s-1)}  \nabla_{(\mu_1} \nabla^{\lambda} 
\left( - \frac{2}{2s-3} \right) 
g_{\mu_2 \mu_3} \nabla^\nu \sigma_{\mu_4 \ldots \mu_{s-1}\lambda) \nu} \cong
\frac{2}{(2s-1)} \nabla_{(\mu_1} \nabla_{\mu_2} \nabla^\lambda \sigma_{\mu_3\ldots\mu_{s-1})\lambda} \ .
\end{equation}
Combining (\ref{a1}), (\ref{a2}) and (\ref{a3}) then leads to 
\begin{eqnarray}\label{Lact}
\bigl({\cal L}^{(s)} \xi^{(\sigma)}  \bigr)_{\mu_1\ldots\mu_{s-1}} 
& \cong & \frac{1}{(2s-1)} \nabla_{(\mu_1}  \Biggl[
4 (s-1) \left( -\Delta + \frac{s^2-s+1}{\ell^2} \right) \, \sigma_{\mu_2\ldots \mu_{s-1})} \nonumber \\
& & \qquad \qquad \qquad \quad + (5-2s) 
\nabla_{\mu_2} \nabla^\lambda \sigma_{\mu_3\ldots\mu_{s-1})\lambda} \Biggr] \ .
\end{eqnarray}
For the simplest case, $s=3$, we have also worked out the trace piece; this is described in 
appendix~\ref{app:B}. 
\medskip

Now the important observation is that the differential operator in the square brackets of (\ref{Lact}) 
acts on $\sigma_{(s-2)}$ in precisely the same way as the differential operator in (\ref{quadvp1}) acts on
$\tilde\vp_{(s-2)}$. Both $\sigma_{(s-2)}$ and $\tilde\vp_{(s-2)}$ are symmetric 
traceless, but not necessarily transverse tensors of rank $(s-2)$, and thus the eigenvalues of 
the two operators 
agree exactly, including multiplicities. As a consequence the contribution to the path integral 
coming from $\sigma_{(s-2)}$ cancels precisely against that arising from integrating out 
$\tilde\vp_{(s-2)}$. 

In particular, the only contributions that actually survive are those coming from
$\vp^{\rm TT}_{(s)}$, see eq.\ (\ref{phitt}), as well as the contribution of $\xi^{\rm TT}_{(s-1)}$ 
to the ghost determinant, see (\ref{xitrans}). Putting these two contributions together gives the full one-loop amplitude for general $s$ in the simple form 
\begin{equation}
Z^{(s)} =  \Biggl[ \det\left(-\Delta  + \frac{s(s-3)}{\ell^2}  \right)^{\rm TT}_{(s)}\Biggr]^{-\frac{1}{2}}  \, 
\Biggl[ \det\left(-\Delta + \frac{s(s-1)}{\ell^2} \right)^{\rm TT}_{(s-1)}\Biggr]^{\frac{1}{2}}  \ ,
\end{equation}
thus proving \eq{dets}.

\section{One loop Determinants and Holomorphic Factorisation}

Given the explicit formula for $Z^{(s)}$, it is now straightforward to calculate the one loop
determinant on thermal AdS$_3$. As was explained in detail in \cite{David:2009xg}, the 
relevant determinant is of the form 
\be{7.1}
- \log {\det \left( - \Delta + \frac{m_s^2}{\ell^2}\right)^{\rm TT}_{(s)} } = \int_0^\infty \frac{dt}{t}  \, K^{(s)}(\tau,\bar{\tau}; t)  \, 
e^{-m_s^2 t}\ , 
\ee
where $K^{(s)}$ is the spin $s$ heat kernel 
\begin{equation}
\label{thermalh3-ker}
K^{(s)}(\tau,\bar{\tau}; t)=\sum_{m=1}^{\infty}{\tau_2\over \sqrt{4\pi t} 
|\sin{m\tau\over 2}|^2}\cos(s m\tau_1)e^{-{m^2\tau_2^2\over 4 t}}e^{-(s+1)t} \ .
\end{equation}
Here $q=e^{i\tau}$, with $\tau=\tau_1+i\tau_2$ the complex structure modulus of the 
$T^2$ boundary of thermal AdS$_3$. 
Note that for the case at hand we have for the helicitiy $s$ component $m_s^2 = s(s-3)$, and 
hence the total $t$-exponent is $s (s-3) +(s+1) = (s-1)^2$, while 
for the helicity $(s-1)$ component $m_{s-1}^2=s(s-1)$ and we get 
$s(s-1) + s = s^2$.
Performing the $t$-integral with the help of the identity  
\begin{equation}
\label{integral}
{1\over \sqrt{4\pi}}\int_0^{\infty}{dt\over t^{3/2}}e^{-{\alpha^2\over 4t}-\beta^2 t}
={1\over \alpha}e^{-\alpha\beta} \ , 
\end{equation}
we therefore obtain 
\begin{eqnarray}
- \log {\det \left( - \Delta + \frac{s(s-3)}{\ell^2}\right)^{\rm TT}_{(s)} } & =  & 
\sum_{m=1}^{\infty} \frac{1}{m}\, {\cos(s m\tau_1)\over |\sin{m\tau\over 2}|^2} \, e^{-m\tau_2 (s-1)} \nonumber \\
& = & \sum_{m=1}^{\infty} \frac{2}{m}\, 
\frac{1}{|1-q^m|^2}
 \left( q^{ms} + \bar{q}^{ms} \right) \ ,
\end{eqnarray}
as well as 
\begin{eqnarray}
- \log {\det \left( - \Delta + \frac{s(s-1)}{\ell^2}\right)^{\rm TT}_{(s-1)} } & =  & 
\sum_{m=1}^{\infty} \frac{1}{m}\, {\cos((s-1) m\tau_1)\over |\sin{m\tau\over 2}|^2} \, e^{-m\tau_2 s} \nonumber \\
& = & \sum_{m=1}^{\infty} \frac{2}{m}\, 
\frac{q^m \bar{q}^m}{|1-q^m|^2}\, \left( q^{m(s-1)} + \bar{q}^{m(s-1)} \right) \ .
\end{eqnarray}
Hence we find for 
\begin{eqnarray}
- \log Z^{(s)} & = & \sum_{m=1}^{\infty} \frac{1}{m}\, 
\frac{1}{|1-q^m|^2} \Bigl[ 
\left(q^{ms} + \bar{q}^{ms}\right) - 
q^m \bar{q}^m \left( q^{m(s-1)} + \bar{q}^{m(s-1)} \right) \Bigr]  \nonumber \\
& = & \sum_{m=1}^{\infty} \frac{1}{m}\, 
\frac{1}{|1-q^m|^2} \Bigl[  q^{ms} (1 - \bar{q}^m) + \bar{q}^{ms} (1 - q^m) \Bigr] \nonumber \\
& = & \sum_{m=1}^{\infty} \frac{1}{m}\,  \Bigl[
\frac{q^{ms}}{1-q^m} + \frac{\bar{q}^{ms}}{1-\bar{q}^m} \Bigr] \ .
\end{eqnarray}
Thus the result is the sum of a purely holomorphic, and a purely anti-holomorphic expression. 
Expanding the denominator by means of a geometric series and noting that the sum over $m$ just
gives the series expansion of the logarithm we hence obtain
\begin{equation}
Z^{(s)} = \prod_{n=s}^{\infty} \frac{1}{|1-q^{n}|^2} \ ,
\end{equation}
thus proving (\ref{detevl}).

\section{${\cal W}_{N}$, ${\cal W}_{\infty}$ and the MacMahon Function}

As is explained in \cite{Campoleoni:2010zq} it is consistent to consider higher spin gauge theories
with only finitely many spin fields. More specficially, the construction of \cite{Campoleoni:2010zq}
in terms of a Chern-Simons action based on $SL(N)\times SL(N)$ leads to a theory that has, 
in addition to the graviton of spin $s=2$, a family of fields of spin $s=3,\ldots, N$. The quadratic 
part of its action is just the sum of the actions $S[\varphi_{(s)}]$ with $s=2,\ldots, N$. The above calculation
therefore implies that the corresponding one-loop determinant equals
\begin{equation}
Z_{SL(N)} = \prod_{s=2}^{N} \prod_{n=s}^{\infty}\, \frac{1}{|1-q^n|^2} = \chi_0({\cal W}_N) \times  \bar{\chi}_0({\cal W}_N) \ ,
\end{equation}
where $\chi_0({\cal W}_N)$ is the vacuum character of the ${\cal W}_N$ algebra
\begin{equation}\label{chi0}
\chi_0({\cal W}_N)  = \prod_{s=2}^{N} \prod_{n=s}^{\infty}\, \frac{1}{(1-q^n)} \ ,
\end{equation}
see {\it e.g.}\ Sec.~6.3.2 of \cite{Bouwknegt:1992wg}. Indeed, by the usual Poincare-Birkhoff-Witt 
theorem (see for example \cite{Watts:1990pd}), a basis for the vacuum
representation of ${\cal W}_N$ is given by
\begin{equation}\label{WNbas}
W^{(N)}_{-n^{(N)}_1} \cdots W^{(N)}_{-n_{l_N}^{(N)}} \, W^{(N-1)}_{-n^{(N-1)}_1} \cdots 
W^{(N-1)}_{-n^{(N-1)}_{l_{N-1}}} \cdots  
 W^{(2)}_{-n^{(2)}_1} \cdots W^{(2)}_{-n^{(2)}_{l_{2}}} \, \Omega \ ,
\end{equation}
where $W^{(K)}_n$ are the modes of the field of conformal dimension $K$, and 
\begin{equation}\label{basmod}
n^{(K)}_1\geq n_2^{(K)} \geq \cdots \geq n_{l_K}^{(K)} \geq K \ .
\end{equation}
Here we have used that $W^{(K)}_n\Omega = 0$ for $n\geq -K+1$ --- this is the reason for
the lower bound in (\ref{basmod}) --- but we have assumed that there are no other null vectors in
the vacuum representation;
this will be the case for generic central charge $c$. We have furthermore denoted the Virasoro modes
by $W^{(2)}_n\equiv L_n$. It is then easy to see that (\ref{chi0}) is just
the counting formula for the basis (\ref{WNbas}). Thus our one loop calculation produces the 
partition function of the ${\cal W}_N$ algebra, as suggested by the analysis of \cite{Campoleoni:2010zq}. 

Maloney and Witten \cite{Maloney:2007ud} have argued that the corresponding answer in the case of pure (super-)gravity 
should be one loop exact. Essentially, the argument is that for the representation of the 
(super-)Virasoro symmetry algebra of the theory corresponding to the vacuum character, the energy 
levels cannot be corrected. One may, in principle, have other states contributing to the partition function. 
However, we know semi-classically ($c\rightarrow \infty$) that there are no propagating gravity states 
in the bulk. Therefore any additional states that might contribute to the partition function must have energies 
going to infinity in the semi-classical limit, such as black hole states or other geometries. 
But these would correspond to non-perturbative corrections from the point of view of the bulk path 
integral computation. 

All the ingredients of this argument are also present in our case of massless higher spin theories. 
We do not have any propagating states in the bulk, and the only semiclassical physical states are 
the generalised Brown-Henneaux excitations of the vacuum. These are boundary states and their 
energies are governed by the ${\cal W}_N$ algebra as argued above. Thus any additional 
contributions would be non-perturbative, and it follows that the above one loop answer is perturbatively 
exact. 

In \cite{Henneaux:2010xg} a classical Brown-Henneaux analysis was also 
performed for the Blencowe theory based on (two copies of)
the infinite dimensional Vasiliev 
higher spin algebra $hs(1,1)$ \cite{Blencowe:1988gj,Bergshoeff:1989ns}. This 
theory possesses one spin field for each spin $s=2,3,\ldots$, and thus the one-loop partition function 
becomes the $N\rightarrow \infty$ limit of $Z_{SL(N)}$, {\it i.e.}\ 
\begin{equation}\label{Zhs}
Z_{hs(1,1)} =  \prod_{s=2}^{\infty} \prod_{n=s}^{\infty}\, \frac{1}{|1-q^n|^2} 
= |M(q)|^{2}\,  \times  \prod_{n=1}^{\infty} |1-q^n|^2 \ ,
\end{equation}
where $M(q)$ is the MacMahon function (\ref{macm}). Note, in particular, that the 
MacMahon function is essentially the ${\cal W}_{\infty}$ vacuum character. This connection 
appears not to be widely known.\footnote{This form of the character for ${\cal W}_{\infty}$ (or rather 
${\cal W}_{1+\infty}$) appears, for instance, in \cite{Awata:1994xm,Awata:1994tf}, and the connection 
also features in the appendix of  \cite{Heckman:2006sk}. We thank B.~Szendroi for bringing these 
references to our attention.}

\section{Final Remarks}

We have seen that a computation of the leading quantum effects for higher spin theories 
on AdS$_3$ can be  carried out explicitly. Our result suggests strongly that the quantum Hilbert 
space can be organised in terms of the vacuum representation of the ${\cal W}_N$ algebra. 
This also leads to the 
conclusion that this answer is perturbatively exact. Thus we have control over the quantum theory, 
at least to all orders in the power series expansion in Newton's constant. It would be very interesting 
to understand whether the full non-perturbative quantum theory is well defined. 
In the case of pure gravity it was argued in \cite{Maloney:2007ud}  that,
under some reasonable looking assumptions, pure gravity on AdS$_3$ does not exist non-perturbatively. 
It would be very interesting to revisit this question for the higher spin theories considered here. In 
particular, one may hope that the situation could be different for the $hs(1,1)$ theory with 
${\cal W}_{\infty}$ symmetry. We are currently investigating this question and 
hope to report on it shortly.
A positive answer would probably give some impetus to investigations 
of these symmetries in higher dimensional AdS spacetimes.\footnote{${\cal W}_{N}$ and 
${\cal W}_{\infty}$ algebras have also appeared as spacetime symmetries of non-critical string theories,
see {\it e.g.}\  \cite{Gava:1991bd,Dhar:1992hr}. It would be interesting to explore the connection, 
if any, to the above realisations.}

In this context we find the appearance of the MacMahon function as the ${\cal W}_{\infty}$ 
vacuum character very significant. The MacMahon function first appeared in string theory in the 
non-perturbative investigation of topological strings \cite{Gopakumar:1998ii,Gopakumar:1998jq}. 
It was further interpreted in terms of a quantum stringy Calabi-Yau geometry in 
\cite{Okounkov:2003sp,Iqbal:2003ds}. Perhaps, we should now interpret the ubiquitous 
appearance of the MacMahon function in the context of topological strings in terms of a hidden 
${\cal W}_{\infty}$ symmetry. It is also rather suggestive that the MacMahon function
(together with the $\eta$-function prefactor of (\ref{Zhs}))  precisely accounts for the contribution 
of the supergravity modes to the elliptic genus of M-theory on 
AdS$_3\times S^2\times X_6$ \cite{Gaiotto:2006ns,Kraus:2006nb}. This might provide a 
concrete link between their appearance in topological strings and in AdS$_3$. 

At a technical level, it might be interesting to redo the analysis of the quadratic fluctuations within the Chern-Simons formulation of the higher spin theories \cite{Blencowe:1988gj}. 
Finally, we should remark that the considerations of this paper 
should be straightforwardly extendable to the 
supersymmetric case. The techniques of \cite{David:2009xg} apply equally well to fermionic fields,
and for example the one loop answer for $s={3\over 2}$ was already explicitly worked out in 
\cite{David:2009xg}, leading to a super-Virasoro character. Thus one might naturally expect to find 
supersymmetric versions of ${\cal W}_{N}$ and ${\cal W}_{\infty}$ vacuum characters for the 
appropriate supersymmetric higher spin theories \cite{Sezgin:1989fa,Pope:1989mg}.

\bigskip

{\bf Acknowledgements:} We would like to specially thank Justin David for collaboration
in the initial stages of this work  as well as for many discussions on related matters. 
We would also like to thank Dileep Jatkar and Ashoke Sen for helpful discussions. 
M.R.G.\ thanks Harvard University and CalTech for
hospitality while this work was being done. His work is also supported 
partially by the Swiss National Science Foundation. 
R.G.'s work was partly supported by a 
Swarnajayanthi Fellowship of the Dept.\ of Science and Technology, Govt.\ of India 
and as always by the support for basic science by the Indian people.

\section*{Appendix} 
\appendix

\section{Conventions}\label{app:A}

In our conventions the commutator of two covariant derivatives, evaluated on a totally symmetric rank $s$
contravariant tensor, is equal to 
\begin{equation}\label{curv}
[ \nabla_{\mu}, \nabla_{\nu}] \, \xi^{\rho_1\ldots \rho_s} = \sum_{j=1}^{s} R^{\rho_j}{}_{\delta\mu\nu}\,  
\xi^{\rho_1 \ldots \widehat{\rho}_j \ldots \rho_s \delta} \ ,
\end{equation}
where the notation $\widehat{\rho}_j$ means that $\rho_j$ is excluded. The Riemann curvature tensor for 
AdS$_3$ is of the form 
\begin{equation}
R_{\mu\nu\rho\sigma} = - \frac{1}{\ell^2} \left( g_{\mu\rho} g_{\nu\sigma} - g_{\mu\sigma} g_{\nu\rho} \right) \ .
\end{equation}
The Ricci tensor is then
\begin{equation}
R^{\mu}{}_{\nu\mu\sigma} = - \frac{2}{\ell^2}\, g_{\nu\sigma} \ .
\end{equation}
We shall also use the conventions of \cite{Campoleoni:2010zq} that 
by an index $(\mu_1\ldots \mu_s)$ we mean the symmetrised expression without
any combinatorial factor, but with the understanding that terms that are obviously symmetric
will not be repeated. So for example, the tensor $\nabla_{(\mu_1} \xi_{\mu_2\ldots \mu_{s})}$ equals
\begin{equation}
\nabla_{(\mu_1} \xi_{\mu_2\ldots \mu_{s})} = 
\sum_{j=1}^{s} \nabla_{\mu_j} \xi_{\mu_1\ldots \widehat{\mu}_j \ldots \mu_{s}} \ ,
\end{equation}
if $\xi_{(s-1)}$ is a symmetric tensor, {\it etc}.  By $\Delta$ we always mean the Laplace operator
\begin{equation}
\Delta = \nabla^\lambda \, \nabla_\lambda \ .
\end{equation}
Because of (\ref{curv}), the explicit action depends on the spin $s$ of the field
on which $\Delta$ acts.

\section{The calculation for $s=3$}\label{app:3de}

In this appendix we give some of the details of the calculation of section~\ref{sec:3}. First we explain 
how to obtain our explicit formula (\ref{Sxi}) for the $\xi$-dependent exponent (\ref{quad1}) of 
(\ref{gh1}). To start with we plug (\ref{decomp1}) into (\ref{quad1}) to obtain 
\begin{eqnarray}
S_\xi & = & 3 \int d^3x \sqrt{g}  \, \Biggl[ 
\xi^{{\rm TT}}_{\nu\rho}\,  \left(- \Delta + \frac{6}{\ell^2} \right) \xi^{{\rm TT}\, \nu\rho} \nonumber \\
& & \qquad \qquad \qquad  
+ \nabla_{(\nu} \sigma^{\rm T}_{\rho)} \,  \left(- \Delta + \frac{6}{\ell^2} \right) \nabla^{(\nu} \sigma^{{\rm T}\rho)} 
- {6\over 5}\,  \nabla_{(\nu} \sigma^{{\rm T}}_{\rho)} \, \nabla^\rho\, \nabla_\mu 
\nabla^{(\mu} \sigma^{{\rm T} {\nu)}}   \nonumber \\
& & \qquad \qquad \qquad  
+ \psi_{\nu\rho} \,  \left(- \Delta + \frac{6}{\ell^2} \right) \psi^{\nu\rho}
- \frac{6}{5}\,  \psi_{\nu\rho} \, \nabla^\rho \nabla_\mu \, 
\psi^{\mu\nu} \Biggr] 
\ .\label{quad2}
\end{eqnarray}
Note that there are no cross-terms between $\xi^{\rm TT}$, $\sigma^{\rm T}$ 
and $\psi^{\mu\nu}$, simply because any potential index contractions lead to vanishing 
results on account of the tracelessness and transversality of $\xi^{\rm TT}$ and 
$\sigma^{{\rm T}\nu}$. We want to simplify the expressions in the second and third line. 

First we consider the $\sigma^{{\rm T}\nu}$ terms. To this end we observe that 
\begin{equation}\label{ob1}
\left(- \Delta + \frac{6}{\ell^2} \right) \nabla_{(\mu} \sigma^{\rm T}_{\nu)} = 
\nabla_{(\mu} \, \left(- \Delta + \frac{10}{\ell^2} \right) \sigma^{\rm T}_{\nu)} \ ,
\end{equation}
as one checks explicitly. It then follows that the first term of the second line of (\ref{quad2}) leads to 
\begin{eqnarray}
S_\xi^{[\sigma,1]} & = & 3 \int d^3x \sqrt{g}  \, \Biggl[ \nabla_{(\nu} \sigma^{\rm T}_{\rho)} \Biggr] \, 
\Biggl[\left(- \Delta + \frac{6}{\ell^2} \right) \nabla^{(\nu} \sigma^{{\rm T}\rho)} \Biggr] \nonumber \\
& =  & 
6 \int d^3x \sqrt{g}  \,  \sigma^{\rm T}_{\nu} \left(-\Delta + \frac{2}{\ell^2} \right) \, 
\left(-\Delta + \frac{10}{\ell^2} \right)   \sigma^{{\rm T}\nu} \ .
\end{eqnarray}
In order to evaluate the second term of the second line we now calculate
\begin{equation}
\nabla^\mu \, \nabla_{(\mu} \sigma^{\rm T}_{\nu)} = 
\Delta \sigma^{\rm T}_{\nu} - \frac{2}{\ell^2} \sigma^{\rm T}_\nu \ ,
\end{equation}
where we have used the transversality of $\sigma^{{\rm T}\mu}$, (\ref{strans}). 
Using integration by parts we therefore get
\begin{eqnarray}
S_\xi^{[\sigma,2]} 
& = & {18\over 5} \int d^3x \sqrt{g}  \, \, \sigma^{\rm T}_\nu \left(-\Delta + \frac{2}{\ell^2} \right) \, 
\left(-\Delta + \frac{2}{\ell^2} \right)  \sigma^{{\rm T}\nu} \ .
\end{eqnarray}
Putting the two calculations together we thus arrive at the result
\begin{equation}\label{fin1}
S_\xi^{[\sigma,1]} + S_\xi^{[\sigma,2]} = {48\over 5} \int d^3x \sqrt{g}  \, \, 
\sigma^{\rm T}_{\nu} \left(-\Delta + \frac{2}{\ell^2} \right) \, 
\left(-\Delta + \frac{7}{\ell^2} \right)  \, \sigma^{{\rm T}\nu} \ ,
\end{equation}
which is the second line of (\ref{Sxi}). 
\smallskip

\noindent 
Next we deal with the $\psi^{\mu\nu}$ terms. The analogue of  (\ref{ob1}) is now 
\begin{equation}
\left(- \Delta + \frac{6}{\ell^2} \right) \psi_{\mu\nu} = 
\left(\nabla_\mu \nabla_\nu - \frac{1}{3} g_{\mu\nu} \Delta \right) \, 
\left( -\Delta + \frac{12}{\ell^2} \right) \psi \ .
\end{equation}
The first term of the third line then leads to 
\begin{eqnarray}
S_\xi^{[\psi,1]} & = & 3 \int d^3x \sqrt{g}  \, 
\Biggl[ \left( \nabla_\nu \nabla_\rho - \frac{1}{3} g_{\mu\rho}\Delta \right) \psi \Biggr] \cdot 
\Biggl[
\left(- \Delta + \frac{6}{\ell^2} \right)  \left( \nabla^\nu \nabla^\rho - \frac{1}{3} g^{\mu\rho} \Delta \right) \psi \Biggr]
\nonumber \\
& =  & 
2\, \int d^3x \sqrt{g}  \,  \, \psi \left( - \Delta \right) \, 
\left(-\Delta + \frac{3}{\ell^2} \right) \, \left(-\Delta + \frac{12}{\ell^2} \right) \psi \ .
\end{eqnarray}
For the second term we calculate
\begin{eqnarray}
\nabla_\rho \nabla^\mu  \psi_{\mu\nu} 
& = & \frac{2}{3} \, \nabla_\rho \nabla_\nu \left( \Delta - \frac{3}{\ell^2} \right) \psi \ , \label{pro}
\end{eqnarray}
thus leading to
\begin{eqnarray}
S_\xi^{[\chi,2]} & = & -\frac{18}{5}  \int d^3x \sqrt{g}  \, 
\Biggl[ \left( \nabla^\nu \nabla^\rho - \frac{1}{3} g^{\nu\rho}\Delta \right) \psi \Biggr] \cdot 
\Biggl[
\left(\nabla_\rho \nabla^\mu  \right)  \left( \nabla_\mu \nabla_\nu - \frac{1}{3} g_{\mu\nu} \Delta \right) \psi \Biggr]
\nonumber \\
& =  & 
 \frac{8}{5}\, \int d^3x \sqrt{g}  \,  \, \psi \left( - \Delta \right) \, 
\left(-\Delta + \frac{3}{\ell^2} \right) \, \left(-\Delta + \frac{3}{\ell^2} \right) \psi \ .
\end{eqnarray}
Putting the two calculations together we therefore get
\begin{equation}\label{fin2}
S_\xi^{[\chi,1]} + S_\xi^{[\chi,2]} = \frac{18}{5}\, \int d^3x \sqrt{g}  \,  \, \psi \left( - \Delta \right) \,  
\left(-\Delta + \frac{3}{\ell^2} \right) \, \left(-\Delta + \frac{8}{\ell^2} \right)
\, \psi \ ,
\end{equation}
which is the third line of (\ref{Sxi}).

\subsection{The full action of ${\cal L}^{(3)}$}\label{app:B}

In this appendix we work out the full action of ${\cal L}^{(3)}$, including the trace piece.
Actually, in order to do this calculation efficiently, it is convenient to modify ${\cal L}^{(3)}$
by a trace piece so that it maps traceless tensors to traceless tensors. The resulting
operator is 
\begin{equation}
\bigl(\hat{\cal L}^{(3)} \xi \bigr)^{\nu\rho} = \left( - \Delta + \frac{6}{\ell^2} \right)  \xi^{\nu\rho}
- \frac{3}{5} \left( \nabla^\nu \nabla_\mu \xi^{\mu\rho} + \nabla^\rho \nabla_\mu \xi^{\mu\nu} \right) 
+ \frac{2}{5} \, g^{\nu\rho} \nabla_\alpha \nabla_\beta \, \xi^{\alpha\beta} \ .
\end{equation}
Next we consider the action of $\hat{\cal L}^{(3)}$ on the traceless tensor
\begin{equation}
\xi^{(\sigma)\, \mu\nu} \equiv \nabla^{\mu} \sigma^{\nu} + \nabla^{\nu} \sigma^{\mu} 
- \frac{2}{3}\, g^{\mu\nu} \, \nabla_\alpha \sigma^\alpha \ .
\end{equation}
After a lengthy calculation one finds 
\begin{equation}\label{L3a}
\bigl(\hat{\cal L}^{(3)} \xi^{(\sigma)} \bigr)^{\nu\rho} =
\frac{1}{5} \Biggl[
8 \,  \nabla^{(\nu} \left( -\Delta + \frac{7}{\ell^2} \right) \sigma^{\rho)} 
-  \nabla^{(\nu} \nabla^{\rho)} \left( \nabla_\lambda \sigma^\lambda\right) 
- 6\,g^{\nu\rho}\, \nabla_\lambda \left(-\Delta + \frac{6}{\ell^2} \right) \sigma^\lambda \Biggr] \ .
\end{equation}
The first two terms obviously agree with (\ref{Lact}) for $s=3$. To understand how to obain
(\ref{L3a}) let us look at the various terms of $\hat{\cal L}^{(3)}$ separately: from the first term of
$\hat{\cal L}^{(3)}$ one gets
\begin{equation}
\left( - \Delta + \frac{6}{\ell^2} \right) \xi^{(\sigma)\, \nu\rho} =  
\nabla^{(\nu} \left( - \Delta + \frac{10}{\ell^2} \right) \sigma^{\rho)}  
+ \frac{2}{3} \, g^{\nu\rho} \, \nabla_\lambda \Delta \sigma^\lambda 
- \frac{20}{3\, \ell^2} \, g^{\nu\rho} \, (\nabla_\lambda \sigma^\lambda) \ . \label{f1}
\end{equation}
One easily checks that the right-hand side is indeed traceless, as must be.
We group the remaining terms into two traceless parts, namely 
\begin{eqnarray}
& & - \frac{3}{5} \left( \nabla^\nu \nabla_\mu \nabla^{(\mu} \sigma^{\rho)} 
+ \nabla^\rho \nabla_\mu \nabla^{(\mu} \sigma^{\rho)}  \right) 
+ \frac{2}{5} g^{\nu\rho}\, \nabla_\lambda \nabla_\nu \nabla^{(\lambda} \sigma^{\nu)} \nonumber \\
& & \qquad = 
 \frac{3}{5} \nabla^{(\nu} \left(-\Delta + \frac{2}{\ell^2} \right) \sigma^{\rho)} 
- \frac{3}{5}  \nabla^{(\nu} \nabla^{\rho)} (\nabla_\lambda \sigma^\lambda) 
+ \frac{4}{5} g^{\nu\rho} \, \nabla_\lambda \Delta\, \sigma^\lambda \ , \label{f2}
\end{eqnarray}
and 
\begin{eqnarray}
& & - \frac{3}{5} \Bigl( \nabla^\nu \nabla_\mu (-{\textstyle \frac{2}{3} })  g^{\mu\rho} 
(\nabla_\lambda \sigma^\lambda)
+ \nabla^\rho \nabla_\mu (-{\textstyle \frac{2}{3} })  g^{\mu\nu} (\nabla_\lambda \sigma^\lambda)\Bigr)
+ \frac{2}{5} g^{\nu\rho}\, \nabla_\lambda \nabla_\tau 
(-{\textstyle \frac{2}{3} })  g^{\lambda\tau} (\nabla_\alpha \sigma^\alpha) \nonumber \\
& &\qquad
= \frac{2}{5} \nabla^{(\nu} \nabla^{\rho)} \left(\nabla_\lambda \sigma^\lambda\right)
-\frac{4}{15} g^{\nu\rho}\,  \nabla_\lambda \left( \Delta +\frac{2}{\ell^2} \right)\, \sigma^\lambda \ .\label{f3}
\end{eqnarray}
One then easily checks that the sum of (\ref{f1}), (\ref{f2}) and (\ref{f3}) gives indeed the right
hand-side of (\ref{L3a}).

\bibliographystyle{JHEP}

\end{document}